\RequirePackage{lineno}
\documentclass[aps,prl,twocolumn,superscriptaddress]{revtex4}

\usepackage{graphicx}
\usepackage{bm}
\usepackage{amsmath}
\usepackage{amssymb}
\usepackage{hyperref}

\begin{document}
\title{Chirality-Dependent Motion Transmission between Aligned Carbon Nanotubes}

\author{Zhao Wang}
\email{zw@gxu.edu.cn}

\affiliation{Department of Physics, Guangxi University, Nanning 530004, China}
\affiliation{Institute of Materials Chemistry, TU Wien, A-1060 Vienna, Austria}

\begin{abstract}
I demonstrate a directional motion-transmission behavior of aligned carbon nanotubes (CNTs) using atomistic simulations. The network of overlapping $\pi$ orbitals at the interface act as gear teeth to translate the sliding motion of a CNT into a rotating motion of the adjacent CNT, or \textit{viceversa}. The efficiency of this orthogonal motion transmission is found to strongly depend on the tube chirality, by which the interfacial stacking configuration of the atoms is determined. These results have strong implications on the design of the motion transmission system at the nanoscale.
\end{abstract}

\maketitle

\section{Introduction}

The miniaturization of devices and machines down to the nanometer scale presents exciting opportunities for various applications, half a century after the prescient prediction by Richard Feynman at an American Physical Society meeting banquet on Dec. 29, 1959 \cite{leong2004}. Taming the mechanical motion of the nanodevices is one of the major challenges in taking the miniaturization technology envisioned by Feynman beyond mundane geometric scaling. For instance, considerable attention has recently been devoted to the ``top-down'' design of nano-machines such as molecular cars \cite{Shirai2006}, motors \cite{Kassem2017}, elevators \cite{Badjic2004} and shuttles\cite{Brouwer2001}. Light can be used as the power input of such molecular machines \cite{ErbasCakmak2015}, however the induced motion is rather random due to the lack of a motion-transmission system, which is the key component in most machines and devices with moving parts. Such a problem remains critical for the design of nano-machines inspired by their macroscopic-world counterparts. The difficulties arising at the nanoscale include not only the synthesis of structures with well positioned cut-teeth, but also the strong friction and adhesion caused by a dramatic increase in the surface-to-volume ratio \cite{Kim2007}.

Carbon nanotubes (CNTs) have received considerable attention with a view to their use as components in nanoscale devices thanks to their unique mechanical properties and peculiar one-dimensional structure \cite{Dong2007}. The possibility of a gear effect has theoretically been proposed for CNTs with benzyne-derived teeth \cite{Han1997}. However, attaching molecules at specific positions on the CNT surface is way beyond the resolution of the present chemical functionalization technology \cite{Pantarotto2008}. Recently, a screw-like motion has been reported for shells in concentric CNTs \cite{Cai2014,Kolmogorov2000a,Bourlon2004}, with the potential to achieve directional molecule transport with low friction \cite{Guerra2017}. Furthermore, the interaction between CNTs and substrates was found to strongly depend on the interface registry \cite{Kolmogorov2004,Falvo2000,Chen2013,Sinclair2018}. Can these surface-interaction features of CNTs be used to design a motion-transmission system without needing precise chemical functionalization? Here we demonstrate a motion-transmission behavior of aligned CNTs by means of molecular mechanics. The chirality-dependent interaction between two aligned CNTs is used to convert the axial sliding of a driving tube into a rotating motion of the adjacent one, or \textit{viceversa}.

\section{Method}

The CNT exhibits a \textit{chiral} structure that can be characterized by a pair of indices $n$ and $m$, which define a helical circumferential vector in the hexagonal carbon lattice \cite{CNTbook2001}. In our simulations, we consider a pair of infinite single-walled CNTs (denoted as CNT$_{1}$ and CNT$_{2}$) that are aligned side by side in the van der Waals (VDW) adhesion. This consideration is important because the CNTs as prepared by chemical vapor deposition (CVD) are often in bundle alignment \cite{Ren1998}. Two experimentally-possible scenarios are considered. In the first case, CNT$_{1}$ is made to slide axially with a constant rate of $0.5\,$nm per $10^{6}$ iterations while the atoms in CNT$_{2}$ are constrained to move in the plane normal to its axis as shown in Fig.~\ref{F1} (a). In the second scheme, CNT$_{1}$ is caused to rotate around its axis with a rate of $2\pi$ per $10^{6}$ iterations, while the CNT$_{2}$ atoms are set to be free in vacuum. The response of CNT$_{2}$, driven by the motion of CNT$_{1}$, is simulated by a molecular mechanics procedure in which a metastable state (a motionless equilibrium atomistic configuration of the atoms) is computed at each iteration by minimizing the total potential energy $\varepsilon$ of the system using a gradient descent algorithm \cite{Wang2007g,Wang2007,Wang2009carbon}. We here use the molecular mechanics instead of the molecular dynamics \cite{Guo2015,Wu2019} in order to avoid the thermal noise. Note that the pressure between the CNTs holds roughly zero since the inter-tube spacing is allowed to change during the simulation in the both schemes. The simulation schemes are demonstrated by the video recordings provided in the Supplemental Materials.

$\varepsilon$ is given by the sum of the pairwise covalent and long-range interactions,

\begin{equation}
\label{Eq0}
\varepsilon=\sum \limits_{i=1}^{N_{1}-1}{\sum\limits_{\substack{j=i+1} }^{N_{1}}{ \varepsilon^{cov}_{ij}}}
                 +\sum \limits_{k=1}^{N_{2}-1}{\sum\limits_{\substack{l=k+1} }^{N_{2}}{ \varepsilon^{cov}_{kl}}}
								 +\sum \limits_{i=1}^{N_{1}}{\sum\limits_{\substack{k=1} }^{N_{2}}{ \varepsilon^{vdw}_{ik}}}
\end{equation}

\noindent where $i$, $j$, $k$ and $l$ are indices of atoms, $N_{1}$ and $N_{2}$ are the total number of atoms in CNT$_{1}$ and CNT$_{2}$, respectively. The first and second terms on the right side of Eq.\ref{Eq0} represent the potential energy of the covalent bonds of the CNT$_{1}$ and the CNT$_{2}$ respectively. They are given by the second generation of reactive empirical bond-order (REBO) force field, in which the total interatomic potential involves many-body terms,

\begin{equation}
\label{Eq1}
\varepsilon^{cov}_{ij}=
\begin{array}{l}
\varphi^R\left(r_{ij}\right)
+b_{ij}\varphi^A\left(r_{ij}\right)
+\sum\limits_{\substack{k=1 \\ k\ne i,j} }^N
{\sum\limits_{\substack{\ell=1 \\ \ell\ne i,j,k}}^N
{\varphi_{kij\ell}^{tor}}}
\end {array}
\end{equation}

\noindent where $\varphi^R$ and $\varphi^A$ denote the interatomic repulsion and attraction terms between the valence electrons, respectively. $\varphi^{tor}$ represents the effect of single-bond torsion. The many-body effects are included in the bond-order function $b_{ij}$, which depends on the atomic distance, the bond angle, the dihedral angle and the bond conjugation. The parameterization and benchmarks for this potential have been provided elsewhere \citep{Brenner2002}. The REBO potential is reported to afford a good description of the structural flexibility of low-dimensional carbons \cite{Wang2009d,Wang2011,Wang2018,Qi2018}.

The Lennard-Jones (LJ) force field is employed to describe the inter-tube interaction potential $\varepsilon^{vdw}$,

\begin{equation}
\label{Eq2}
\varepsilon^{vdw}_{ik} =  4 \epsilon \left[ \left( \frac{\sigma}{r_{ik}}  \right)^{12} -  \left( \frac{\sigma}{r_{ik}}  \right)^{6}      \right]
\end{equation}

\noindent with a potential well depth of $\epsilon=2.4\;\mathrm{meV}$, an equilibrium distance of $\sigma=0.34\;\mathrm{nm}$ and a cutoff radius of $1.2\;\mathrm{nm}$ \citep{Stuart2000a}. The LJ potential has been reported to underestimate the surface energy corrugation \cite{Kolmogorov2005,Sinclair2018}. We therefore compared the results obtained by using the LJ force field with those based on the more sophisticated Kolmogorov-Crespi (KC) force field, which has been reported to give an improved description to the overlap of $\pi$-orbitals between two graphene layers at a high load \cite{Kolmogorov2005}. The comparison, however, reveals only minor differences between the KC and LJ models in the context of our simulations, as shown in the Supplemental Materials. This may be due to the fact that our CNTs are placed in vacuum and the interfacial pressure is therefore about zero. Hence, the inter-tube force strongly depends on the position of the peaks of the interaction potential surface, while the depth of the potential well has a far weaker effect.

\section{Results and Discussion}

\begin{figure}[htp]
\centerline{\includegraphics[width=8cm]{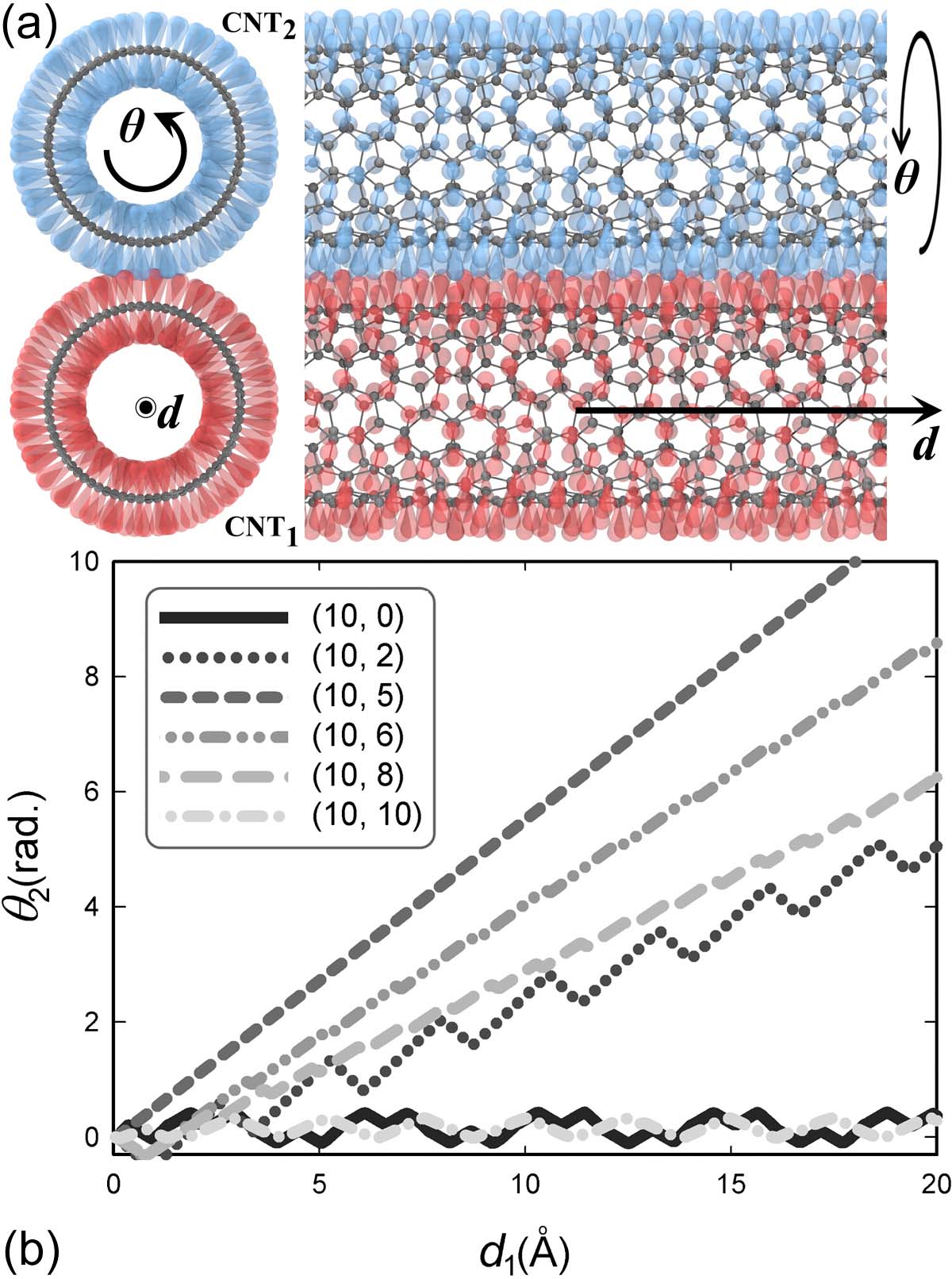}}
\caption{\label{F1}
(a) Model setup for a pair of aligned (10,5) CNTs in the first simulation scheme (\textit{left}: side view, \textit{right}: top view). The gray spheres stand for the carbon atoms. The overlapping of round cones illustrates the way how the atoms in a CNT interact with those in another tube. The rotation of the driven CNT$_{2}$ is found to follow a right-hand rule with the thumb pointing to the sliding direction of the driving CNT$_{1}$. (b) Rotation angle $\theta$ of the CNT$_{2}$ \textit{versus} displacement $d$ of the CNT$_{1}$ for (10, $m$) tubes with $m=0,1,2,...,10$.}
\end{figure}

We first consider the simple case of two aligned identical CNTs in the first simulation scheme. A spontaneous rotation of CNT$_{2}$ is observed when CNT$_{1}$ is caused to slide along its axis in the direction shown in Fig.~\ref{F1} (a). Let us denote the rotation angle of the CNT$_{2}$ by $\theta_{2}$, and plot its values in Fig.~\ref{F1} (b) \textit{versus} the sliding distance ($d_{1}$) of CNT$_{1}$. A directional motion transmission behavior is observed. Roughly, $\theta_{2}$ is proportional to $d_{1}$ with different proportionality constants for different CNT pairs. For instance, a (10,5) tube is found to rotate the most, with $\theta_{2}$ being a linear function of $d_{1}$. In contrast, oscillations can be found in the $\theta-d$ curves for other tubes. In particular, the armchair (10,10) and zigzag (10,0) CNTs rotate back and forth with a period in the $d_{1}$ variable of about $2.82\;$\AA\;and $4.12\;$\AA, respectively. These periodic lengths coincide with the dimensions of an orthogonal unit cell of the CNT \cite{CNTbook2001}.

\begin{figure}[htp]
\centerline{\includegraphics[width=8cm]{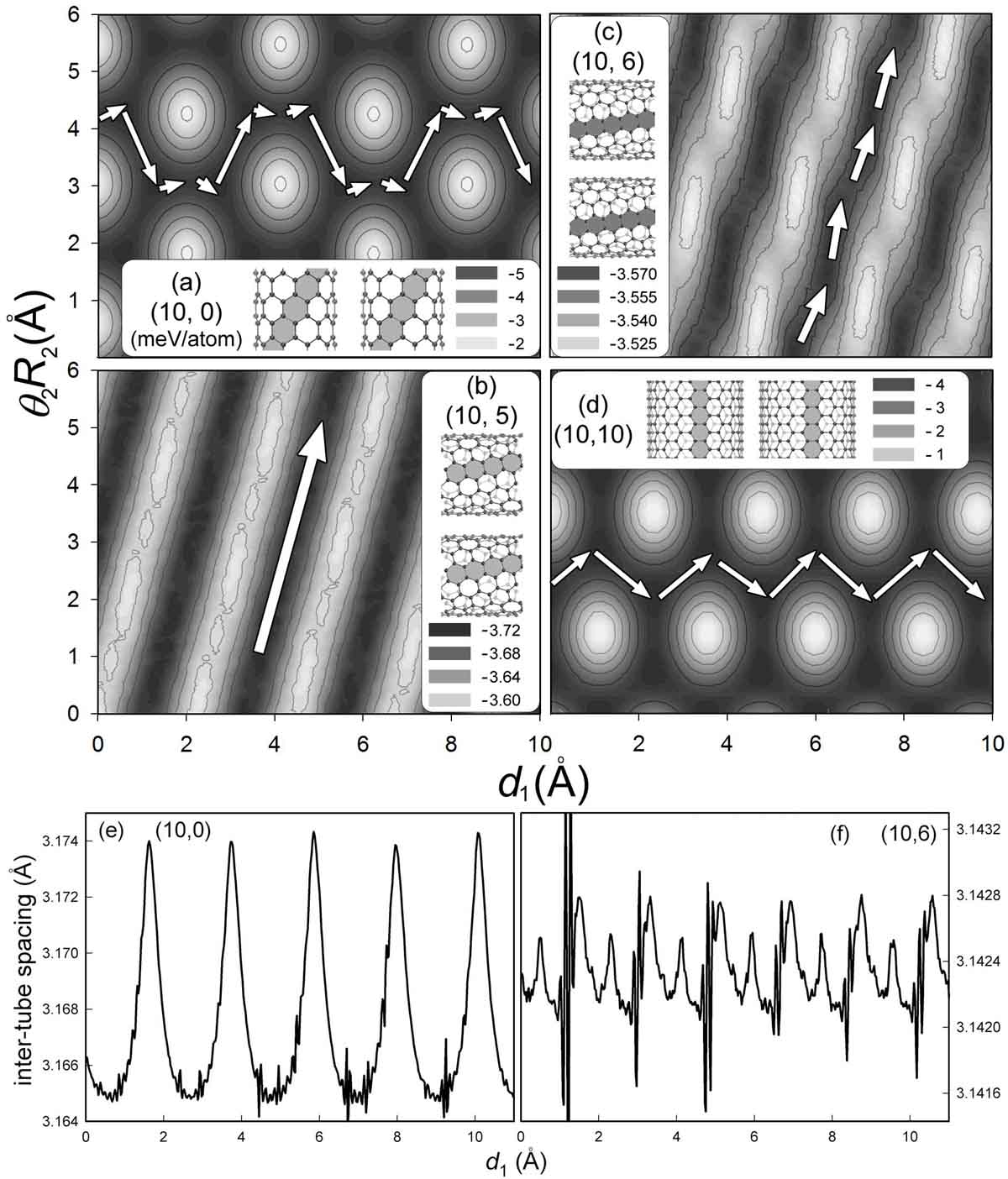}}
\caption{\label{F2}
(a-d) Distribution of the potential energy (DPE)  on the CNT surface for different CNT pairs with respect to $d_{1}$ and $R_{2}\theta_{2}$. Here $R_{2}$ is the radius of CNT$_{2}$. The data are obtained by displacing CNT$_{2}$ atop CNT$_{1}$ in both longitudinal and transverse (angular) directions keeping a minimum inter-tube spacing of $3.0\;$\AA, while the inter-tube distance is allowed to change during the simulations for all other results presented in this paper. The gray scale corresponds to the interaction energy per unit of tube length (meV/\AA). The arrows point along the energetically favorable paths. Insets show how the atomic structure of a CNT is correlated to its chirality. (e-f)  Inter-tube spacing \textit{versus} $d_{1}$ for the (10,0) and (10,6) CNT pairs.}
\end{figure}

Why is there such motion transmission between the aligned CNTs, and why does it show a strong dependence on the chirality? The CNTs exhibit quasi-$sp^{2}$ hybridization, wherein each carbon atom builds three in-plane $\sigma$ bonds, leaving a $\pi$-orbital pointing in the direction normal to the surface. This hybridization forms an atomic hill-and-valley potential landscape like the one shown in Fig.~\ref{F2}(a). When the CNTs are physically adsorbed together in van der Waals interaction, there will be some directions in which the CNTs can move over each other more easily than in other directions \cite{Dienwiebel2004}. These optimized directions follow the principle to minimize the energy corrugation and strongly depend on the structure of CNTs. It has been shown in experiments that the interactions between CNTs and/or graphene sheets mainly depend on the registry of the electron bonds at the interface \cite{Chen2013,Falvo2000}, which is determined by the CNT chirality \cite{Kolmogorov2004,Guerra2017}. To progress towards a detailed explanation of this effect, we depict the distribution of the potential energy (DPE) of the inter-tube interaction for four different pairs of CNTs in Fig.~\ref{F2}. On the DPE, the rotation of CNT$_{2}$ is represented by the change in the ordinate axis (where it has been multiplied by the tube radius $R$ to express it as a length), while the sliding of CNT$_{1}$ is given by the variation in the abscissa axis. Without taking into account finite-temperature corrections, the change in $\theta_{2}$ in response to a slight variation of $d_{1}$ should always follow a minimum-energy trajectory. This forces the interface between two tubes to ``surf'' the waves of the DPE along a path corresponding to the lowest energy corrugation. Such energy-optimized paths (EOPs) are indicated by the arrows in Fig.~\ref{F2}. 

The motion-transmission behavior of the aligned CNTs shown in Fig.~\ref{F1} (b) can be explained by the EOP in the DPE landscape, since the shape and periodic length of these EOPs are consistent with those of the $\theta$-$d$ curves in Fig.~\ref{F1} (b). For instance, we see that the shape of the EOP of the (10,6) tubes is sinusoidal [Fig.~\ref{F2} (c)], while that of the (10,5) tubes is almost a straight line [Fig.~\ref{F2} (d)]. These coincide with the shape of the $\theta$-$d$ curves of the corresponding tubes in Fig.~\ref{F1} (b). We also observe the zigzag- and armchair-shaped EOPs in Figs.\ref{F2} (a) and (d) for the (10,0) and (10,10) tubes, respectively. Note that $\theta_{2}R_{2}$ is here used instead of $\theta_{2}$  for making a valid comparison between CNTs of different radii. 

Moreover, we computed the inter-tube spacing as a function of the sliding distance for two different tube pairs as shown in Fig.~\ref{F2} (e-f). The oscillation of the data is relatively small due to the fact that there is no temperature fluctuation and the tube is free to rotate in our simulations. The oscillation is smaller for the spacing between the (10,6) CNTs since they have shallower energy corrugation as shown in Fig.~\ref{F2} (c). Benchmark density functional theory (DFT) calculations are performed for computing the potential energy of the interaction between CNTs versus the inter-tube distance as shown in the Supplemental Materials. It is seen that the equilibrium minimum distance is different for CNT pairs of different chirality. It ranges from $3.03$ to $3.20$ \AA. The distance data in the Fig.~\ref{F2} (e-f) calculated from classical simulations fall into this range. The minimum distance between two CNTs is usually smaller than that between two graphene layers (approximately $3.4$ \AA) in graphite due to the effect of surface curvature. These results are roughly consistent with previous measurements of experiments \cite{Thess1996} and DFT calculations \cite{Charlier1995, Reich2002}. 

\begin{figure}[htp]
\centerline{\includegraphics[width=8cm]{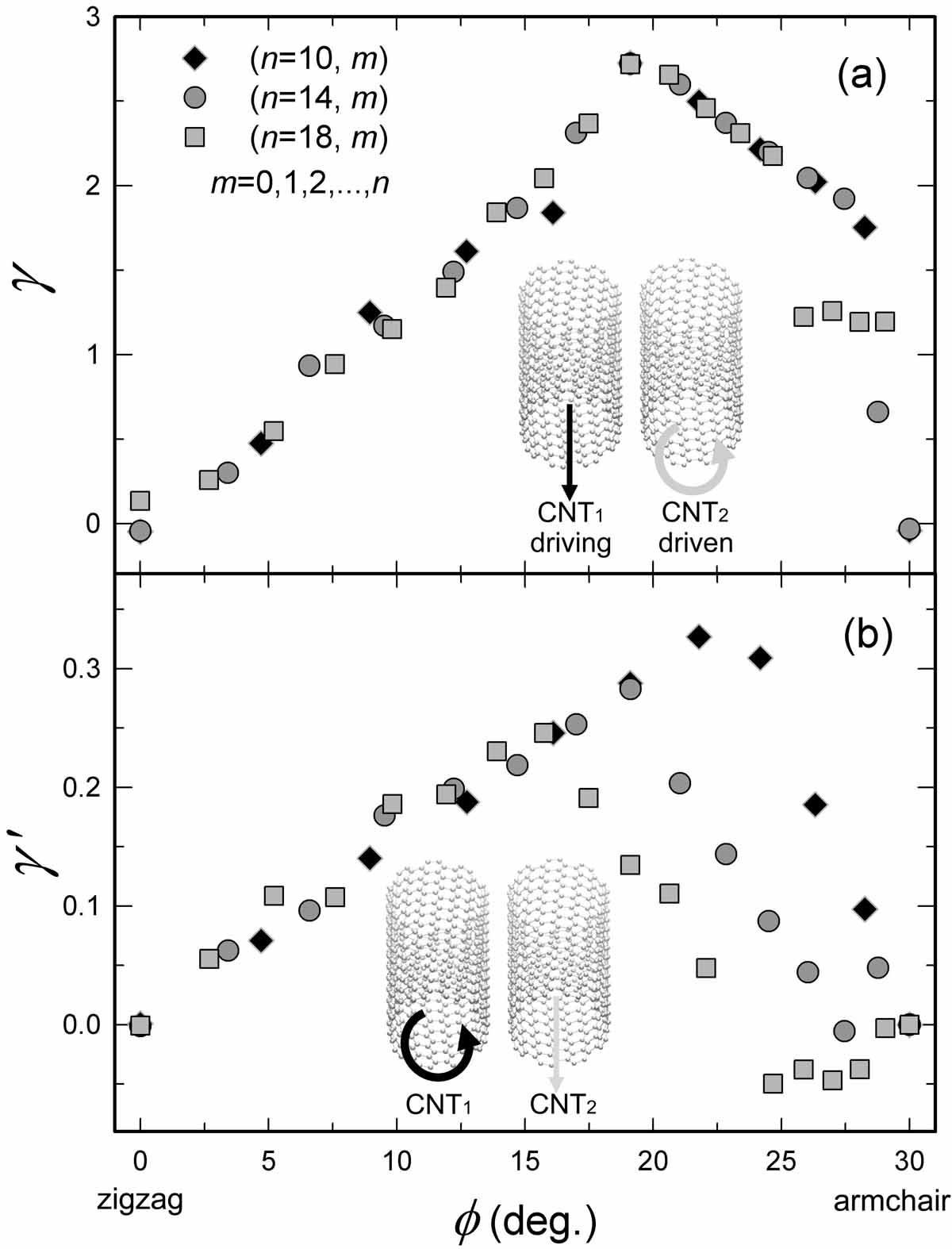}}
\caption{\label{F3}
(a) motion-transmission factor $\gamma$ (Eq.\ref{Eq3}) \textit{vs.} the chiral angle $\phi$ for three different sets of ($n, m$) CNT pairs with $n = 10, 14, 18$ and $m = 0,1,2,...,n$. The inset shows that the driving CNT$_{1}$ is controlled to slide and the CNT$_{2}$ rotates in spontaneous response. (b) Another motion-transmission factor $\gamma'$ (Eq.\ref{Eq4}) for a different case in which the CNT$_{1}$ is controlled to rotate.
}
\end{figure}

Fig.~\ref{F1}(b) shows that $\theta_{2}$ is roughly a linear function of $d_{1}$. We can therefore define a dimensionless factor $\gamma$ that represents the efficiency of motion transmission between two aligned CNTs,

\begin{equation}
\label{Eq3}
\gamma = \frac{R_{2}\theta_{2}}{d_{1}}.
\end{equation}

To further quantify the chiral effect, in Fig.~\ref{F3} (a) we plot $\gamma$ as a function of the chiral angle $\phi$, which is the angle between the chiral vector and the zigzag direction, and is equal to $\arctan \left(\sqrt{3}m/(m+2n)\right)$. It can be seen that $\gamma$ is almost zero for the zigzag ($n,0$) tubes ($\phi=0$) since such a tube keeps oscillating instead of rotating. $\gamma$ increases linearly with increasing $\phi$ before reaching a maximum at $\phi \approx 19.1^{\circ}$, namely the ``magic'' angle of CNTs \cite{Rao2012}. $\gamma$ then decreases to zero for the armchair ($n,n$) tubes beyond this threshold. This trend holds for the three different sets of CNT pairs of ($n, m$) with $n = 10, 14, 18$ and $m = 0,1,2,...,n$. Note that $\gamma$ can be greater than $1.0$.

The rotation of CNTs has been realized in experiments \cite{Cohen-Karni2006}. The motion-transmission behavior discussed above of two aligned CNTs should be reversible since they could function like a rack with a pinion. i.e., the rotation of one of the tubes should be able to induce a sliding motion of the other one. This is confirmed by our simulations of the second scheme, in which the CNT$_{1}$ is made to rotate axially. Another transmission factor can be assigned to such a case as follows,

\begin{equation}
\label{Eq4}
\gamma' = \frac{d_{2}}{R_{1}\theta_{1}}
\end{equation}

\noindent where $R_{1}$ and $\theta_{1}$ are the radius and the rotation angle of the driving CNT$_{1}$, respectively, and $d_{2}$ is the sliding distance of the driven CNT$_{2}$. $\gamma'$ is plotted in Fig.~\ref{F3} (b) for pairs of identical CNTs. We see that $\gamma'$ increases linearly with increasing $\phi$ before reaching a critical chiral angle, and that it then decreases rapidly with increasing $\phi$ beyond this threshold. The critical chiral angle is not the same for the three sets of CNT pairs of different sizes, unlike the trend for $\gamma$. This difference may be due to different displacement rate and the fact that CNT$_{2}$ is restricted not to move in the axial direction when measuring $\gamma$ but it is left free for measuring $\gamma'$.

\begin{figure}[htp]
\centerline{\includegraphics[width=8cm]{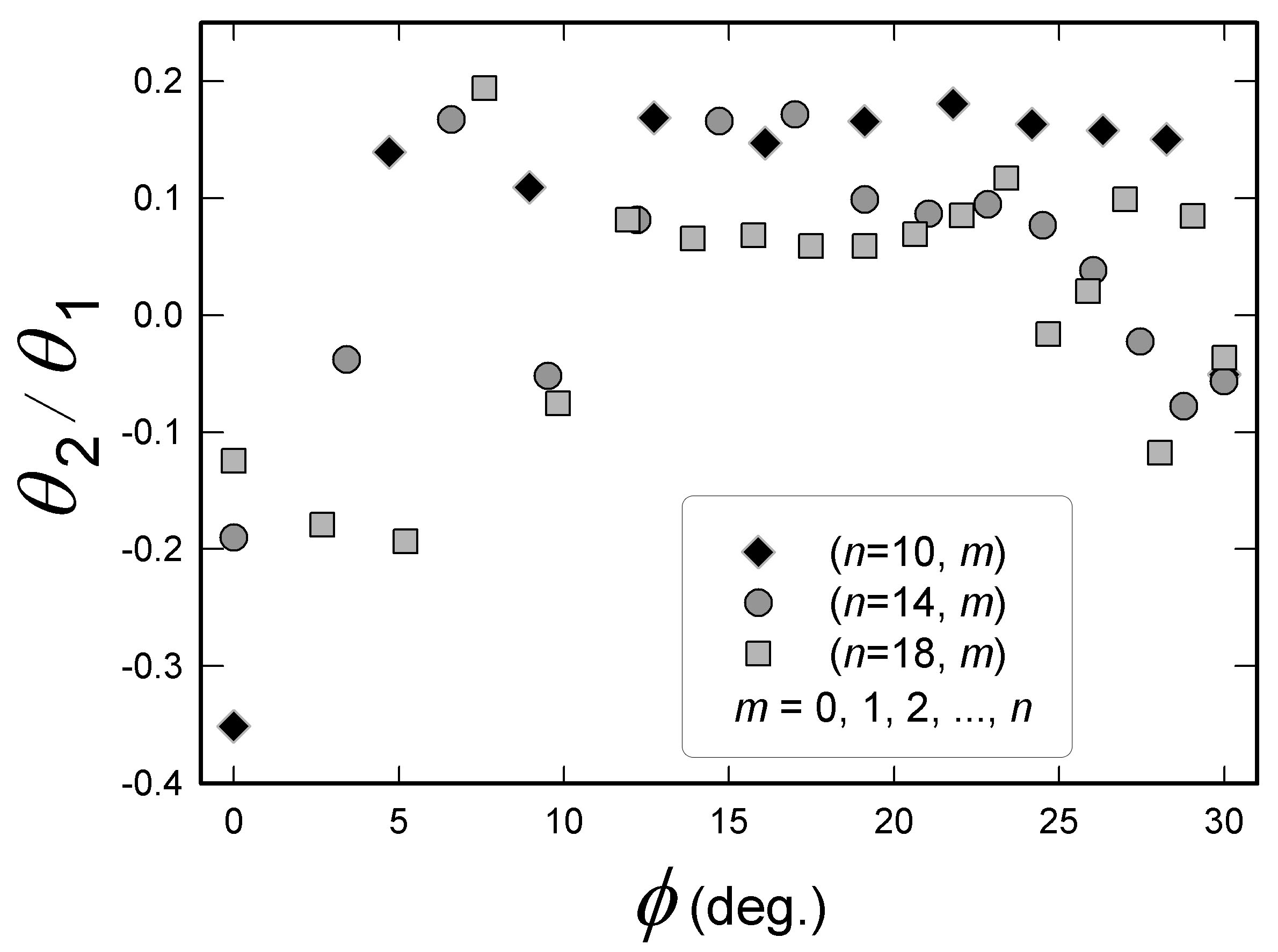}}
\caption{\label{F4}
$\theta_{2}/\theta_{1}$ \textit{versus} the chiral angle $\phi$ in the second simulation scheme for three different sets of ($n, m$) CNT pairs with $n = 10, 14, 18$ and $m = 0,1,2,...,n$.}
\end{figure}

The CNT$_{2}$ is observed to axially rotate when it is driven to slide by the rotation of  CNT$_{1}$ in the second simulation scheme. Its rotation angle $\theta_{2}$ also depends on the CNT chirality as shown in Fig.~\ref{F4}. The ratio $\theta_{2}/\theta_{1}$ can be either positive and negative. The CNT$_{2}$ rotates in the same direction as the CNT$_{1}$ does in case of positive $\theta_{2}/\theta_{1}$, while it rotates in the inverse direction in case of negative $\theta_{2}/\theta_{1}$. It can be seen that $\theta_{2}/\theta_{1}$ is negative for zigzag CNTs, and it increases to positive with increasing chiral angle until $\phi>10^{\circ}$. It then roughly holds constant before decreasing back to negative values for armchair CNTs with increasing $\phi$. The CNT radius also shows influence on the rotation of CNT$_{2}$.

\begin{figure}[htp]
\centerline{\includegraphics[width=8cm]{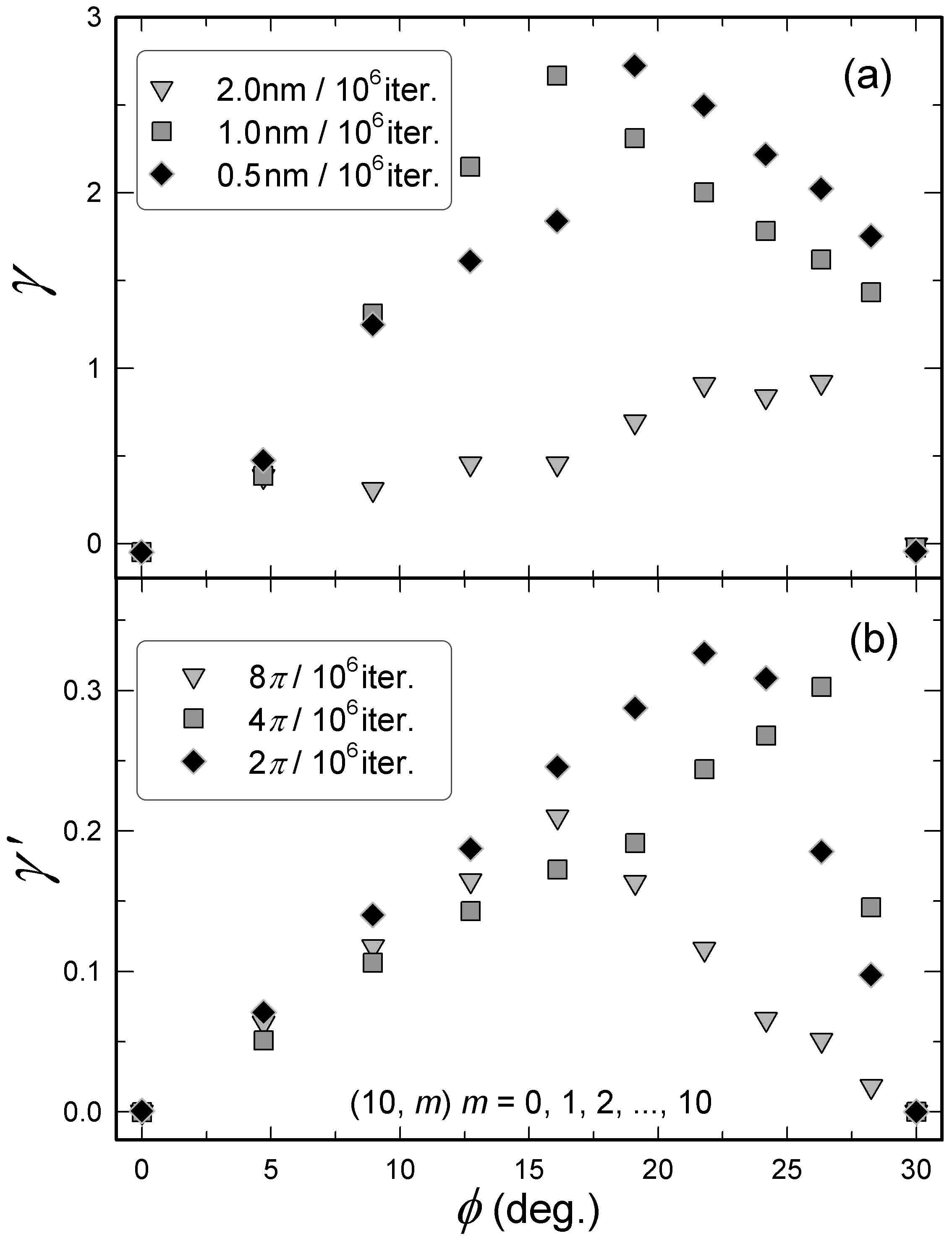}}
\caption{\label{F5}
Motion-transmission factor $\gamma$ (Eq.\ref{Eq3}, scenario 1) (a) and $\gamma'$  (Eq.\ref{Eq4}, scenario 2) (b) \textit{versus} the chiral angle $\phi$ for sets of simulations with different displacement rates.}
\end{figure}

Sets of simulations are also carried out with different displacement rates for the both simulation scenarios. Fig.~\ref{F5} shows the influence of the displacement rate on the motion transmission behavior of CNTs. It is seen that the motion transmission factor is lower when the driving tube moves faster in the first simulation scheme. The position of the peak of $\gamma$ varies with different displacement rates. For the second simulation scheme, similar trend of the transmission factor $\gamma’$ holds for $\phi<16.0^{\circ}$, it then changes with different rotating rate of the driving CNTs. These variations of the transmission factors qualitatively show the important influence of the displacement rate, although quantitative correlation between the transmission factors and the displacement rate cannot be accurately predicted as the time is not well defined by the present simulation method unlike that in molecular dynamics.

\begin{figure}[htp]
\centerline{\includegraphics[width=8cm]{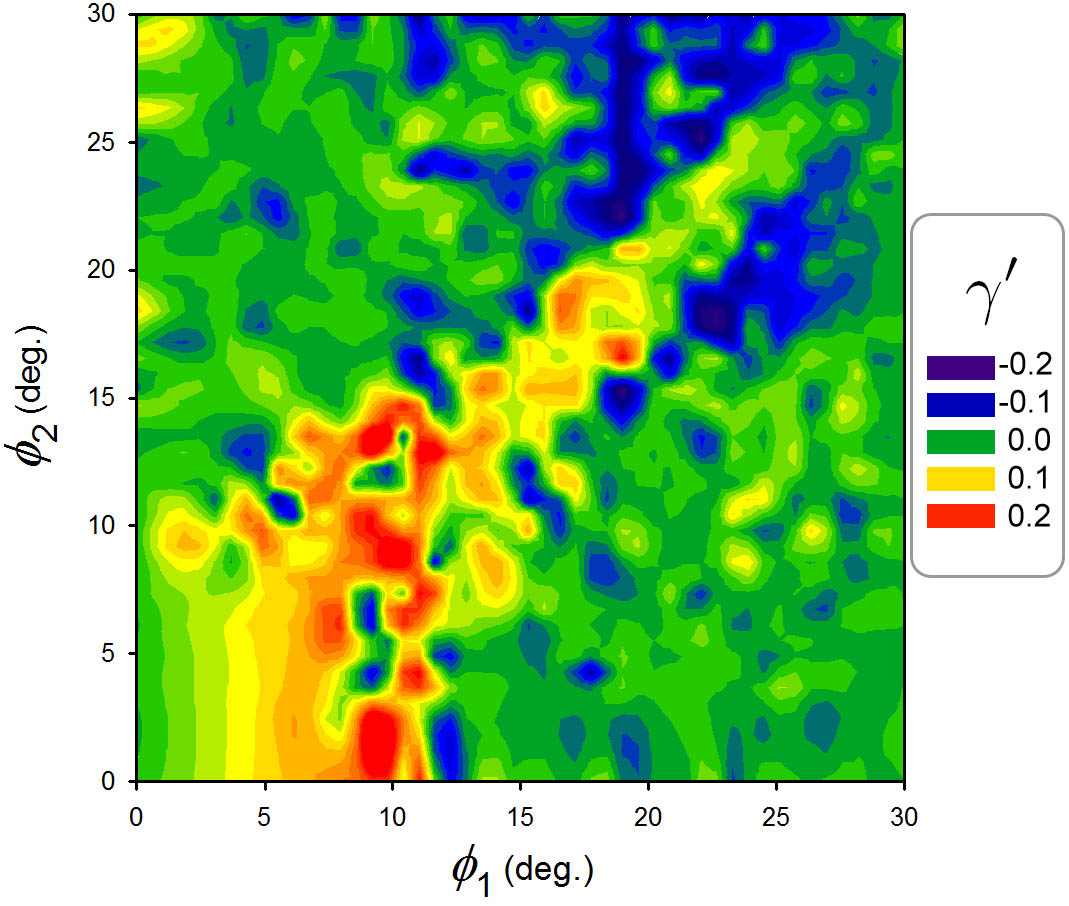}}
\caption{\label{F6}
The motion-transmission ratio $\gamma'$ as a function of the chiral angle of both tubes. The data are from simulations with $900$ pairs of CNTs with different combinations of chiralities. A complete list of these bi-chiral CNT pairs is provided in the Supplemental Materials.}
\end{figure}

The aforementioned results are for identical CNTs. Does motion transmission exist between CNTs of different chirality? To answer this question, we have tested $900$ pairs of bi-chiral CNTs in the second simulation scheme. The measured transmission factor $\gamma'$ is plotted in Fig.~\ref{F6} as a function of the chiral angles of both tubes. It is seen that $\gamma'$ is higher in absolute value for tube pairs of similar chirality, except in the armchair and zigzag cases. We also see that the axial sliding of CNT$_{2}$ can happen in both 'forward' or 'backward' directions depending on the bi-chirality. The asymmetry of the $\gamma'$ landscape may be due to a small strain $<0.01$ that is applied to the bi-chiral tubes with different lattice constants in order to keep the periodic boundary condition. The second scheme is chosen here since it has less artificially imposed boundary condition and considered to be more general than the first scheme. Moreover, we have simulated a CNT bundle composed of $91$ aligned (10,5) CNTs with a central driving tube which is caused to either slide along or rotate around its axis. The later is found to transmit motion: the surrounding tubes slide axially together when the central tube rotates, as shows in the Supplemental Materials. However, when a sliding is applied to the central tube, no rotation of the surrounding tubes is seen. Such a contrast to the two-tube setup is because the rotation motion is ``locked'' between the nearby CNTs.

Finally, we note that the effect of temperature is not involved in the present work using molecular mechanics simulations. However, the thermal effect will add uncertainty to the movement of the CNTs, since the distribution of the potential energy on a surface will become ``fuzzy’’ when the temperature rises. In a previous work of Zhang et al. using molecular dynamics simulations, it was shown that the torque transmission efficiency between walls of concentric CNTs with confined water molecules decreases at increasing temperature \cite{Zhang2012}. Similar effects can be expected on the motion transmission between aligned CNTs. Furthermore, since the driven CNT is free to move in the plane normal to its axis, the pressure between two CNTs is always about zero. The system studied in the present work can be used to transmit motion but can barely be used to transmit force and torque, since the inter-tube distance is adjusted spontaneously. To transmit force and torque, the inter-tube spacing needs to be fixed with external pressure applied between the CNTs. In this scenario, small and multi-walled CNTs may work better than large and single-walled ones owing to possible deformation of the CNT cross section under applied pressure.

\section{Conclusion}
In conclusion, we have demonstrated an orthogonal motion transmission between aligned CNTs. The sliding motion of a CNT can be spontaneously converted to a rotation of the neighboring ones, or \textit{viceversa}. The motion-transmission factors are found to be well defined functions of the tube chiral angle. Analyses on the potential energy distribution show that this chirality dependence is attributable to the stacking of the helical lattice determining the overlap of the network of electron orbitals at the interface. This motion transmission behavior takes advantage of the strong interfacial adhesion between nanostructures, and is expected to be superlubric in nature \cite{Dienwiebel2004}. It thus has important implications to the design of mechanical power transmission systems using nanomaterials with extreme surface-to-volume ratio.


\begin{thebibliography}{39}
\expandafter\ifx\csname natexlab\endcsname\relax\def\natexlab#1{#1}\fi
\expandafter\ifx\csname bibnamefont\endcsname\relax
  \def\bibnamefont#1{#1}\fi
\expandafter\ifx\csname bibfnamefont\endcsname\relax
  \def\bibfnamefont#1{#1}\fi
\expandafter\ifx\csname citenamefont\endcsname\relax
  \def\citenamefont#1{#1}\fi
\expandafter\ifx\csname url\endcsname\relax
  \def\url#1{\texttt{#1}}\fi
\expandafter\ifx\csname urlprefix\endcsname\relax\def\urlprefix{URL }\fi
\providecommand{\bibinfo}[2]{#2}
\providecommand{\eprint}[2][]{\url{#2}}

\bibitem[{\citenamefont{Ieong et~al.}(2004)\citenamefont{Ieong, Doris,
  Kedzierski, Rim, and Yang}}]{leong2004}
\bibinfo{author}{\bibfnamefont{M.}~\bibnamefont{Ieong}},
  \bibinfo{author}{\bibfnamefont{B.}~\bibnamefont{Doris}},
  \bibinfo{author}{\bibfnamefont{J.}~\bibnamefont{Kedzierski}},
  \bibinfo{author}{\bibfnamefont{K.}~\bibnamefont{Rim}}, \bibnamefont{and}
  \bibinfo{author}{\bibfnamefont{M.}~\bibnamefont{Yang}},
  \bibinfo{journal}{Science} \textbf{\bibinfo{volume}{306}},
  \bibinfo{pages}{2057} (\bibinfo{year}{2004}).

\bibitem[{\citenamefont{Shirai et~al.}(2006)\citenamefont{Shirai, Morin,
  Sasaki, Guerrero, and Tour}}]{Shirai2006}
\bibinfo{author}{\bibfnamefont{Y.}~\bibnamefont{Shirai}},
  \bibinfo{author}{\bibfnamefont{J.-F.} \bibnamefont{Morin}},
  \bibinfo{author}{\bibfnamefont{T.}~\bibnamefont{Sasaki}},
  \bibinfo{author}{\bibfnamefont{J.~M.} \bibnamefont{Guerrero}},
  \bibnamefont{and} \bibinfo{author}{\bibfnamefont{J.~M.} \bibnamefont{Tour}},
  \bibinfo{journal}{Chem. Soc. Rev.} \textbf{\bibinfo{volume}{35}},
  \bibinfo{pages}{1043} (\bibinfo{year}{2006}).

\bibitem[{\citenamefont{Kassem et~al.}(2017)\citenamefont{Kassem, van Leeuwen,
  Lubbe, Wilson, Feringa, and Leigh}}]{Kassem2017}
\bibinfo{author}{\bibfnamefont{S.}~\bibnamefont{Kassem}},
  \bibinfo{author}{\bibfnamefont{T.}~\bibnamefont{van Leeuwen}},
  \bibinfo{author}{\bibfnamefont{A.~S.} \bibnamefont{Lubbe}},
  \bibinfo{author}{\bibfnamefont{M.~R.} \bibnamefont{Wilson}},
  \bibinfo{author}{\bibfnamefont{B.~L.} \bibnamefont{Feringa}},
  \bibnamefont{and} \bibinfo{author}{\bibfnamefont{D.~A.} \bibnamefont{Leigh}},
  \bibinfo{journal}{Chem. Soc. Rev.} \textbf{\bibinfo{volume}{46}},
  \bibinfo{pages}{2592} (\bibinfo{year}{2017}).

\bibitem[{\citenamefont{Badjic et~al.}(2004)\citenamefont{Badjic, Balzani,
  Credi, Silvi, and Stoddart}}]{Badjic2004}
\bibinfo{author}{\bibfnamefont{J.~D.} \bibnamefont{Badjic}},
  \bibinfo{author}{\bibfnamefont{V.}~\bibnamefont{Balzani}},
  \bibinfo{author}{\bibfnamefont{A.}~\bibnamefont{Credi}},
  \bibinfo{author}{\bibfnamefont{S.}~\bibnamefont{Silvi}}, \bibnamefont{and}
  \bibinfo{author}{\bibfnamefont{J.~F.} \bibnamefont{Stoddart}},
  \bibinfo{journal}{Science} \textbf{\bibinfo{volume}{303}},
  \bibinfo{pages}{1845} (\bibinfo{year}{2004}).

\bibitem[{\citenamefont{Brouwer et~al.}(2001)\citenamefont{Brouwer, Frochot,
  Gatti, Leigh, Mottier, Paolucci, Roffia, and Wurpel}}]{Brouwer2001}
\bibinfo{author}{\bibfnamefont{A.~M.} \bibnamefont{Brouwer}},
  \bibinfo{author}{\bibfnamefont{C.}~\bibnamefont{Frochot}},
  \bibinfo{author}{\bibfnamefont{F.~G.} \bibnamefont{Gatti}},
  \bibinfo{author}{\bibfnamefont{D.~A.} \bibnamefont{Leigh}},
  \bibinfo{author}{\bibfnamefont{L.}~\bibnamefont{Mottier}},
  \bibinfo{author}{\bibfnamefont{F.}~\bibnamefont{Paolucci}},
  \bibinfo{author}{\bibfnamefont{S.}~\bibnamefont{Roffia}}, \bibnamefont{and}
  \bibinfo{author}{\bibfnamefont{G.~W.~H.} \bibnamefont{Wurpel}},
  \bibinfo{journal}{Science} \textbf{\bibinfo{volume}{291}},
  \bibinfo{pages}{2124} (\bibinfo{year}{2001}).

\bibitem[{\citenamefont{Erbas-Cakmak et~al.}(2015)\citenamefont{Erbas-Cakmak,
  Leigh, McTernan, and Nussbaumer}}]{ErbasCakmak2015}
\bibinfo{author}{\bibfnamefont{S.}~\bibnamefont{Erbas-Cakmak}},
  \bibinfo{author}{\bibfnamefont{D.~A.} \bibnamefont{Leigh}},
  \bibinfo{author}{\bibfnamefont{C.~T.} \bibnamefont{McTernan}},
  \bibnamefont{and} \bibinfo{author}{\bibfnamefont{A.~L.}
  \bibnamefont{Nussbaumer}}, \bibinfo{journal}{Chem. Rev.}
  \textbf{\bibinfo{volume}{115}}, \bibinfo{pages}{10081}
  (\bibinfo{year}{2015}).

\bibitem[{\citenamefont{Kim et~al.}(2007)\citenamefont{Kim, Asay, and
  Dugger}}]{Kim2007}
\bibinfo{author}{\bibfnamefont{S.~H.} \bibnamefont{Kim}},
  \bibinfo{author}{\bibfnamefont{D.~B.} \bibnamefont{Asay}}, \bibnamefont{and}
  \bibinfo{author}{\bibfnamefont{M.~T.} \bibnamefont{Dugger}},
  \bibinfo{journal}{Nano Today} \textbf{\bibinfo{volume}{2}},
  \bibinfo{pages}{22} (\bibinfo{year}{2007}).

\bibitem[{\citenamefont{Dong et~al.}(2007)\citenamefont{Dong, Subramanian, and
  Nelson}}]{Dong2007}
\bibinfo{author}{\bibfnamefont{L.~X.} \bibnamefont{Dong}},
  \bibinfo{author}{\bibfnamefont{A.}~\bibnamefont{Subramanian}},
  \bibnamefont{and} \bibinfo{author}{\bibfnamefont{B.~J.}
  \bibnamefont{Nelson}}, \bibinfo{journal}{Nano Today}
  \textbf{\bibinfo{volume}{2}}, \bibinfo{pages}{12} (\bibinfo{year}{2007}).

\bibitem[{\citenamefont{Han et~al.}(1997)\citenamefont{Han, Globus, Jaffe, and
  Deardorff}}]{Han1997}
\bibinfo{author}{\bibfnamefont{J.}~\bibnamefont{Han}},
  \bibinfo{author}{\bibfnamefont{A.}~\bibnamefont{Globus}},
  \bibinfo{author}{\bibfnamefont{R.}~\bibnamefont{Jaffe}}, \bibnamefont{and}
  \bibinfo{author}{\bibfnamefont{G.}~\bibnamefont{Deardorff}},
  \bibinfo{journal}{Nanotechnology} \textbf{\bibinfo{volume}{8}},
  \bibinfo{pages}{95} (\bibinfo{year}{1997}).

\bibitem[{\citenamefont{Pantarotto et~al.}(2008)\citenamefont{Pantarotto,
  Browne, and Feringa}}]{Pantarotto2008}
\bibinfo{author}{\bibfnamefont{D.}~\bibnamefont{Pantarotto}},
  \bibinfo{author}{\bibfnamefont{W.~R.} \bibnamefont{Browne}},
  \bibnamefont{and} \bibinfo{author}{\bibfnamefont{B.~L.}
  \bibnamefont{Feringa}}, \bibinfo{journal}{Chem. Comm.}
  \textbf{\bibinfo{volume}{13}}, \bibinfo{pages}{1533} (\bibinfo{year}{2008}).

\bibitem[{\citenamefont{Cai et~al.}(2014)\citenamefont{Cai, Yin, Qin, and
  Li}}]{Cai2014}
\bibinfo{author}{\bibfnamefont{K.}~\bibnamefont{Cai}},
  \bibinfo{author}{\bibfnamefont{H.}~\bibnamefont{Yin}},
  \bibinfo{author}{\bibfnamefont{Q.~H.} \bibnamefont{Qin}}, \bibnamefont{and}
  \bibinfo{author}{\bibfnamefont{Y.}~\bibnamefont{Li}}, \bibinfo{journal}{Nano
  Lett.} \textbf{\bibinfo{volume}{14}}, \bibinfo{pages}{2558}
  (\bibinfo{year}{2014}).

\bibitem[{\citenamefont{Kolmogorov and Crespi}(2000)}]{Kolmogorov2000a}
\bibinfo{author}{\bibfnamefont{A.~N.} \bibnamefont{Kolmogorov}}
  \bibnamefont{and} \bibinfo{author}{\bibfnamefont{V.~H.}
  \bibnamefont{Crespi}}, \bibinfo{journal}{Phys. Rev. Lett.}
  \textbf{\bibinfo{volume}{85}}, \bibinfo{pages}{4727} (\bibinfo{year}{2000}).

\bibitem[{\citenamefont{Bourlon et~al.}(2004)\citenamefont{Bourlon, Glattli,
  Miko, Forro, and Bachtold}}]{Bourlon2004}
\bibinfo{author}{\bibfnamefont{B.}~\bibnamefont{Bourlon}},
  \bibinfo{author}{\bibfnamefont{D.~C.} \bibnamefont{Glattli}},
  \bibinfo{author}{\bibfnamefont{C.}~\bibnamefont{Miko}},
  \bibinfo{author}{\bibfnamefont{L.}~\bibnamefont{Forro}}, \bibnamefont{and}
  \bibinfo{author}{\bibfnamefont{A.}~\bibnamefont{Bachtold}},
  \bibinfo{journal}{Nano Lett.} \textbf{\bibinfo{volume}{4}},
  \bibinfo{pages}{709} (\bibinfo{year}{2004}).

\bibitem[{\citenamefont{Guerra et~al.}(2017)\citenamefont{Guerra, Leven,
  Vanossi, Hod, and Tosatti}}]{Guerra2017}
\bibinfo{author}{\bibfnamefont{R.}~\bibnamefont{Guerra}},
  \bibinfo{author}{\bibfnamefont{I.}~\bibnamefont{Leven}},
  \bibinfo{author}{\bibfnamefont{A.}~\bibnamefont{Vanossi}},
  \bibinfo{author}{\bibfnamefont{O.}~\bibnamefont{Hod}}, \bibnamefont{and}
  \bibinfo{author}{\bibfnamefont{E.}~\bibnamefont{Tosatti}},
  \bibinfo{journal}{Nano Lett.} \textbf{\bibinfo{volume}{17}},
  \bibinfo{pages}{5321} (\bibinfo{year}{2017}).

\bibitem[{\citenamefont{Kolmogorov et~al.}(2004)\citenamefont{Kolmogorov,
  Crespi, Schleier-Smith, and Ellenbogen}}]{Kolmogorov2004}
\bibinfo{author}{\bibfnamefont{A.~N.} \bibnamefont{Kolmogorov}},
  \bibinfo{author}{\bibfnamefont{V.~H.} \bibnamefont{Crespi}},
  \bibinfo{author}{\bibfnamefont{M.~H.} \bibnamefont{Schleier-Smith}},
  \bibnamefont{and} \bibinfo{author}{\bibfnamefont{J.~C.}
  \bibnamefont{Ellenbogen}}, \bibinfo{journal}{Phys. Rev. Lett.}
  \textbf{\bibinfo{volume}{92}}, \bibinfo{pages}{085503}
  (\bibinfo{year}{2004}).

\bibitem[{\citenamefont{Falvo et~al.}(2000)\citenamefont{Falvo, Steele, Taylor,
  and Superfine}}]{Falvo2000}
\bibinfo{author}{\bibfnamefont{M.~R.} \bibnamefont{Falvo}},
  \bibinfo{author}{\bibfnamefont{J.}~\bibnamefont{Steele}},
  \bibinfo{author}{\bibfnamefont{R.~M.} \bibnamefont{Taylor}},
  \bibnamefont{and}
  \bibinfo{author}{\bibfnamefont{R.}~\bibnamefont{Superfine}},
  \bibinfo{journal}{Phys. Rev. B} \textbf{\bibinfo{volume}{62}},
  \bibinfo{pages}{10665} (\bibinfo{year}{2000}).

\bibitem[{\citenamefont{Chen et~al.}(2013)\citenamefont{Chen, Shen, Xu, Hu, Xu,
  Wang, Guo, Zhang, Peng, Ding et~al.}}]{Chen2013}
\bibinfo{author}{\bibfnamefont{Y.~B.} \bibnamefont{Chen}},
  \bibinfo{author}{\bibfnamefont{Z.~Y.} \bibnamefont{Shen}},
  \bibinfo{author}{\bibfnamefont{Z.~W.} \bibnamefont{Xu}},
  \bibinfo{author}{\bibfnamefont{Y.}~\bibnamefont{Hu}},
  \bibinfo{author}{\bibfnamefont{H.~T.} \bibnamefont{Xu}},
  \bibinfo{author}{\bibfnamefont{S.}~\bibnamefont{Wang}},
  \bibinfo{author}{\bibfnamefont{X.~L.} \bibnamefont{Guo}},
  \bibinfo{author}{\bibfnamefont{Y.~F.} \bibnamefont{Zhang}},
  \bibinfo{author}{\bibfnamefont{L.~M.} \bibnamefont{Peng}},
  \bibinfo{author}{\bibfnamefont{F.}~\bibnamefont{Ding}}, \bibnamefont{et~al.},
  \bibinfo{journal}{Nature Comm.} \textbf{\bibinfo{volume}{4}},
  \bibinfo{pages}{2205} (\bibinfo{year}{2013}).

\bibitem[{\citenamefont{Sinclair et~al.}(2018)\citenamefont{Sinclair, Suter,
  and Coveney}}]{Sinclair2018}
\bibinfo{author}{\bibfnamefont{R.~C.} \bibnamefont{Sinclair}},
  \bibinfo{author}{\bibfnamefont{J.~L.} \bibnamefont{Suter}}, \bibnamefont{and}
  \bibinfo{author}{\bibfnamefont{P.~V.} \bibnamefont{Coveney}},
  \bibinfo{journal}{Adv. Mater.} \textbf{\bibinfo{volume}{30}},
  \bibinfo{pages}{1705791} (\bibinfo{year}{2018}).

\bibitem[{\citenamefont{Dresselhaus et~al.}(2001)\citenamefont{Dresselhaus,
  Dresselhaus, and Avouris}}]{CNTbook2001}
\bibinfo{editor}{\bibfnamefont{M.~S.} \bibnamefont{Dresselhaus}},
  \bibinfo{editor}{\bibfnamefont{G.}~\bibnamefont{Dresselhaus}},
  \bibnamefont{and} \bibinfo{editor}{\bibfnamefont{P.}~\bibnamefont{Avouris}},
  eds., \emph{\bibinfo{title}{Carbon Nanotubes Synthesis, Structure,
  Properties, and Applications.}} (\bibinfo{publisher}{Springer-Verlag Berlin
  Heidelberg}, \bibinfo{year}{2001}).

\bibitem[{\citenamefont{Ren et~al.}(1998)\citenamefont{Ren, Huang, Xu, Wang,
  Bush, Siegal, and Provencio}}]{Ren1998}
\bibinfo{author}{\bibfnamefont{Z.~F.} \bibnamefont{Ren}},
  \bibinfo{author}{\bibfnamefont{Z.~P.} \bibnamefont{Huang}},
  \bibinfo{author}{\bibfnamefont{J.~W.} \bibnamefont{Xu}},
  \bibinfo{author}{\bibfnamefont{J.~H.} \bibnamefont{Wang}},
  \bibinfo{author}{\bibfnamefont{P.}~\bibnamefont{Bush}},
  \bibinfo{author}{\bibfnamefont{M.~P.} \bibnamefont{Siegal}},
  \bibnamefont{and} \bibinfo{author}{\bibfnamefont{P.~N.}
  \bibnamefont{Provencio}}, \bibinfo{journal}{Science}
  \textbf{\bibinfo{volume}{282}}, \bibinfo{pages}{1105} (\bibinfo{year}{1998}).

\bibitem[{\citenamefont{Wang and Devel}(2007)}]{Wang2007g}
\bibinfo{author}{\bibfnamefont{Z.}~\bibnamefont{Wang}} \bibnamefont{and}
  \bibinfo{author}{\bibfnamefont{M.}~\bibnamefont{Devel}},
  \bibinfo{journal}{Phys. Rev. B} \textbf{\bibinfo{volume}{76}},
  \bibinfo{pages}{195434} (\bibinfo{year}{2007}).

\bibitem[{\citenamefont{Wang et~al.}(2007)\citenamefont{Wang, Devel, Langlet,
  and Dulmet}}]{Wang2007}
\bibinfo{author}{\bibfnamefont{Z.}~\bibnamefont{Wang}},
  \bibinfo{author}{\bibfnamefont{M.}~\bibnamefont{Devel}},
  \bibinfo{author}{\bibfnamefont{R.}~\bibnamefont{Langlet}}, \bibnamefont{and}
  \bibinfo{author}{\bibfnamefont{B.}~\bibnamefont{Dulmet}},
  \bibinfo{journal}{Phys. Rev. B} \textbf{\bibinfo{volume}{75}},
  \bibinfo{pages}{205414} (\bibinfo{year}{2007}).

\bibitem[{\citenamefont{Wang}(2009)}]{Wang2009carbon}
\bibinfo{author}{\bibfnamefont{Z.}~\bibnamefont{Wang}},
  \bibinfo{journal}{Carbon} \textbf{\bibinfo{volume}{47}},
  \bibinfo{pages}{3050} (\bibinfo{year}{2009}).

\bibitem[{\citenamefont{Guo et~al.}(2015)\citenamefont{Guo, Wang, and
  Li}}]{Guo2015}
\bibinfo{author}{\bibfnamefont{W.}~\bibnamefont{Guo}},
  \bibinfo{author}{\bibfnamefont{Z.}~\bibnamefont{Wang}}, \bibnamefont{and}
  \bibinfo{author}{\bibfnamefont{J.}~\bibnamefont{Li}}, \bibinfo{journal}{Nano
  Lett.} \textbf{\bibinfo{volume}{15}}, \bibinfo{pages}{6582}
  (\bibinfo{year}{2015}).

\bibitem[{\citenamefont{Wu et~al.}(2019)\citenamefont{Wu, Yang, and
  Wang}}]{Wu2019}
\bibinfo{author}{\bibfnamefont{Z.~Y.} \bibnamefont{Wu}},
  \bibinfo{author}{\bibfnamefont{X.~L.} \bibnamefont{Yang}}, \bibnamefont{and}
  \bibinfo{author}{\bibfnamefont{Z.}~\bibnamefont{Wang}},
  \bibinfo{journal}{Nanotechnology} \textbf{\bibinfo{volume}{30}},
  \bibinfo{pages}{245601} (\bibinfo{year}{2019}).

\bibitem[{\citenamefont{Brenner et~al.}(2002)\citenamefont{Brenner, Shenderova,
  Harrison, Stuart, Ni, and Sinnott}}]{Brenner2002}
\bibinfo{author}{\bibfnamefont{D.}~\bibnamefont{Brenner}},
  \bibinfo{author}{\bibfnamefont{O.}~\bibnamefont{Shenderova}},
  \bibinfo{author}{\bibfnamefont{J.}~\bibnamefont{Harrison}},
  \bibinfo{author}{\bibfnamefont{S.}~\bibnamefont{Stuart}},
  \bibinfo{author}{\bibfnamefont{B.}~\bibnamefont{Ni}}, \bibnamefont{and}
  \bibinfo{author}{\bibfnamefont{S.}~\bibnamefont{Sinnott}},
  \bibinfo{journal}{J. Phys.: Condens. Matter} \textbf{\bibinfo{volume}{14}},
  \bibinfo{pages}{783} (\bibinfo{year}{2002}).

\bibitem[{\citenamefont{Wang and Philippe}(2009)}]{Wang2009d}
\bibinfo{author}{\bibfnamefont{Z.}~\bibnamefont{Wang}} \bibnamefont{and}
  \bibinfo{author}{\bibfnamefont{L.}~\bibnamefont{Philippe}},
  \bibinfo{journal}{Phys. Rev. Lett.} \textbf{\bibinfo{volume}{102}},
  \bibinfo{pages}{215501} (\bibinfo{year}{2009}).

\bibitem[{\citenamefont{Wang and Devel}(2011)}]{Wang2011}
\bibinfo{author}{\bibfnamefont{Z.}~\bibnamefont{Wang}} \bibnamefont{and}
  \bibinfo{author}{\bibfnamefont{M.}~\bibnamefont{Devel}},
  \bibinfo{journal}{Phys. Rev. B} \textbf{\bibinfo{volume}{83}},
  \bibinfo{pages}{125422} (\bibinfo{year}{2011}).

\bibitem[{\citenamefont{Wang}(2018)}]{Wang2018}
\bibinfo{author}{\bibfnamefont{Z.}~\bibnamefont{Wang}}, \bibinfo{journal}{J.
  Phys. D: Appl. Phys.} \textbf{\bibinfo{volume}{51}}, \bibinfo{pages}{435301}
  (\bibinfo{year}{2018}).

\bibitem[{\citenamefont{Qi et~al.}(2018)\citenamefont{Qi, Picaud, Devel, Liang,
  and Wang}}]{Qi2018}
\bibinfo{author}{\bibfnamefont{H.}~\bibnamefont{Qi}},
  \bibinfo{author}{\bibfnamefont{S.}~\bibnamefont{Picaud}},
  \bibinfo{author}{\bibfnamefont{M.}~\bibnamefont{Devel}},
  \bibinfo{author}{\bibfnamefont{E.}~\bibnamefont{Liang}}, \bibnamefont{and}
  \bibinfo{author}{\bibfnamefont{Z.}~\bibnamefont{Wang}},
  \bibinfo{journal}{APJ} \textbf{\bibinfo{volume}{867}}, \bibinfo{pages}{133}
  (\bibinfo{year}{2018}).

\bibitem[{\citenamefont{Stuart et~al.}(2000)\citenamefont{Stuart, Tutein, and
  Harrison}}]{Stuart2000a}
\bibinfo{author}{\bibfnamefont{S.}~\bibnamefont{Stuart}},
  \bibinfo{author}{\bibfnamefont{A.}~\bibnamefont{Tutein}}, \bibnamefont{and}
  \bibinfo{author}{\bibfnamefont{J.}~\bibnamefont{Harrison}},
  \bibinfo{journal}{J. Chem. Phys.} \textbf{\bibinfo{volume}{112}},
  \bibinfo{pages}{6472} (\bibinfo{year}{2000}).

\bibitem[{\citenamefont{Kolmogorov and Crespi}(2005)}]{Kolmogorov2005}
\bibinfo{author}{\bibfnamefont{A.~N.} \bibnamefont{Kolmogorov}}
  \bibnamefont{and} \bibinfo{author}{\bibfnamefont{V.~H.}
  \bibnamefont{Crespi}}, \bibinfo{journal}{Phys. Rev. B}
  \textbf{\bibinfo{volume}{71}}, \bibinfo{pages}{235415}
  (\bibinfo{year}{2005}).

\bibitem[{\citenamefont{Dienwiebel et~al.}(2004)\citenamefont{Dienwiebel,
  Verhoeven, Pradeep, Frenken, Heimberg, and Zandbergen}}]{Dienwiebel2004}
\bibinfo{author}{\bibfnamefont{M.}~\bibnamefont{Dienwiebel}},
  \bibinfo{author}{\bibfnamefont{G.~S.} \bibnamefont{Verhoeven}},
  \bibinfo{author}{\bibfnamefont{N.}~\bibnamefont{Pradeep}},
  \bibinfo{author}{\bibfnamefont{J.~W.~M.} \bibnamefont{Frenken}},
  \bibinfo{author}{\bibfnamefont{J.~A.} \bibnamefont{Heimberg}},
  \bibnamefont{and} \bibinfo{author}{\bibfnamefont{H.~W.}
  \bibnamefont{Zandbergen}}, \bibinfo{journal}{Phys. Rev. Lett.}
  \textbf{\bibinfo{volume}{92}}, \bibinfo{pages}{126101}
  (\bibinfo{year}{2004}).

\bibitem[{\citenamefont{Thess et~al.}(1996)\citenamefont{Thess, Lee, Nikolaev,
  Dai, Petit, Robert, Xu, Lee, Kim, Rinzler et~al.}}]{Thess1996}
\bibinfo{author}{\bibfnamefont{A.}~\bibnamefont{Thess}},
  \bibinfo{author}{\bibfnamefont{R.}~\bibnamefont{Lee}},
  \bibinfo{author}{\bibfnamefont{P.}~\bibnamefont{Nikolaev}},
  \bibinfo{author}{\bibfnamefont{H.~J.} \bibnamefont{Dai}},
  \bibinfo{author}{\bibfnamefont{P.}~\bibnamefont{Petit}},
  \bibinfo{author}{\bibfnamefont{J.}~\bibnamefont{Robert}},
  \bibinfo{author}{\bibfnamefont{C.~H.} \bibnamefont{Xu}},
  \bibinfo{author}{\bibfnamefont{Y.~H.} \bibnamefont{Lee}},
  \bibinfo{author}{\bibfnamefont{S.~G.} \bibnamefont{Kim}},
  \bibinfo{author}{\bibfnamefont{A.~G.} \bibnamefont{Rinzler}},
  \bibnamefont{et~al.}, \bibinfo{journal}{Science}
  \textbf{\bibinfo{volume}{273}}, \bibinfo{pages}{483} (\bibinfo{year}{1996}).

\bibitem[{\citenamefont{Charlier et~al.}(1995)\citenamefont{Charlier, Gonze,
  and Michenaud}}]{Charlier1995}
\bibinfo{author}{\bibfnamefont{J.~C.} \bibnamefont{Charlier}},
  \bibinfo{author}{\bibfnamefont{X.}~\bibnamefont{Gonze}}, \bibnamefont{and}
  \bibinfo{author}{\bibfnamefont{J.~P.} \bibnamefont{Michenaud}},
  \bibinfo{journal}{EPL} \textbf{\bibinfo{volume}{29}}, \bibinfo{pages}{43}
  (\bibinfo{year}{1995}).

\bibitem[{\citenamefont{Reich et~al.}(2002)\citenamefont{Reich, Thomsen, and
  Ordejon}}]{Reich2002}
\bibinfo{author}{\bibfnamefont{S.}~\bibnamefont{Reich}},
  \bibinfo{author}{\bibfnamefont{C.}~\bibnamefont{Thomsen}}, \bibnamefont{and}
  \bibinfo{author}{\bibfnamefont{P.}~\bibnamefont{Ordejon}},
  \bibinfo{journal}{Phys. Rev. B} \textbf{\bibinfo{volume}{65}},
  \bibinfo{pages}{155411} (\bibinfo{year}{2002}).

\bibitem[{\citenamefont{Rao et~al.}(2012)\citenamefont{Rao, Liptak, Cherukuri,
  Yakobson, and Maruyama}}]{Rao2012}
\bibinfo{author}{\bibfnamefont{R.}~\bibnamefont{Rao}},
  \bibinfo{author}{\bibfnamefont{D.}~\bibnamefont{Liptak}},
  \bibinfo{author}{\bibfnamefont{T.}~\bibnamefont{Cherukuri}},
  \bibinfo{author}{\bibfnamefont{B.~I.} \bibnamefont{Yakobson}},
  \bibnamefont{and} \bibinfo{author}{\bibfnamefont{B.}~\bibnamefont{Maruyama}},
  \bibinfo{journal}{Nature Mater.} \textbf{\bibinfo{volume}{11}},
  \bibinfo{pages}{213} (\bibinfo{year}{2012}).

\bibitem[{\citenamefont{Cohen-Karni et~al.}(2006)\citenamefont{Cohen-Karni,
  Segev, Srur-Lavi, Cohen, and Joselevich}}]{Cohen-Karni2006}
\bibinfo{author}{\bibfnamefont{T.}~\bibnamefont{Cohen-Karni}},
  \bibinfo{author}{\bibfnamefont{L.}~\bibnamefont{Segev}},
  \bibinfo{author}{\bibfnamefont{O.}~\bibnamefont{Srur-Lavi}},
  \bibinfo{author}{\bibfnamefont{S.~R.} \bibnamefont{Cohen}}, \bibnamefont{and}
  \bibinfo{author}{\bibfnamefont{E.}~\bibnamefont{Joselevich}},
  \bibinfo{journal}{Nature Nanotech.} \textbf{\bibinfo{volume}{1}},
  \bibinfo{pages}{36} (\bibinfo{year}{2006}).

\bibitem[{\citenamefont{Zhang et~al.}(2012)\citenamefont{Zhang, Ye, Liu, Ding,
  Cheng, Ling, Zheng, Wang, and Wang}}]{Zhang2012}
\bibinfo{author}{\bibfnamefont{Z.~Q.} \bibnamefont{Zhang}},
  \bibinfo{author}{\bibfnamefont{H.~F.} \bibnamefont{Ye}},
  \bibinfo{author}{\bibfnamefont{Z.}~\bibnamefont{Liu}},
  \bibinfo{author}{\bibfnamefont{J.~N.} \bibnamefont{Ding}},
  \bibinfo{author}{\bibfnamefont{G.~G.} \bibnamefont{Cheng}},
  \bibinfo{author}{\bibfnamefont{Z.~Y.} \bibnamefont{Ling}},
  \bibinfo{author}{\bibfnamefont{Y.~G.} \bibnamefont{Zheng}},
  \bibinfo{author}{\bibfnamefont{L.}~\bibnamefont{Wang}}, \bibnamefont{and}
  \bibinfo{author}{\bibfnamefont{J.~B.} \bibnamefont{Wang}},
  \bibinfo{journal}{J. Appl. Phys.} \textbf{\bibinfo{volume}{111}},
  \bibinfo{pages}{114304} (\bibinfo{year}{2012}).

\end{thebibliography}

\end{document}